# Comparing N-Node Set Importance Representative results with Node Importance Representative results for Categorical Clustering: An exploratory study

H.Venkateswara Reddy[1], Dr.S.Viswanadha Raju[2], B.Ramasubba Reddy[3],

[1] Depatment of CSE, Vardhaman College of Engineering, Hyderabad, India
venkat_nidhish@yahoo.co.in
[2] Depatment of CSE, JNTUH College of Engineering, Karim Nagar, India
viswanadha_raju2004@yahoo.com
[3] Depatment of CSE, Ramappa College of Engineering, Warangal, India
b. ramasubbareddy@gmail.com

**Abstract.** The proportionate increase in the size of the data with increase in space implies that clustering a very large data set becomes difficult and is a time consuming process. Sampling is one important technique to scale down the size of dataset and to improve the efficiency of clustering. After sampling, allocating unlabeled objects into proper clusters is impossible in the categorical domain. To address the problem, Chen employed a method called MARDL to allocate each unlabeled data point to the appropriate cluster based on NIR and NNIR algorithms. This paper took off from Chen's investigation and analyzed and compared the results of NIR and NNIR, leading to the conclusion that the two processes contradict each other when it comes to finding the resemblance between an unlabeled data point and a cluster. A new and better way of solving the problem was arrived at that finds resemblance between unlabeled data point within all clusters, while also providing maximal resemblance for allocation of data in the required cluster.

# 1  Introduction

Clustering is an important technique in data mining to partition a dataset into several groups so that data points within the same group are very similar to each other than to data points in other groups, according to some predefined similarity measurements [1]-[5]. The similar groups are labeled clusters. Clustering finds application in manufacturing, medicine, machine learning, information retrieval and research and development planning [6, 7]. Clustering very large datasets is difficult and time consuming [8]-[11]. Sampling is therefore employed to scale down the size of the database to improve the efficiency of clustering [12]-[15]. In sampling, the clustering chosen is a randomly small set of data points (from the given data set) which are used in a clustering algorithm on the sampled data set, which is generally small. The clustering result thus obtained from the sampled data set is expected to be similar to the clustering result of original data set. This makes for an efficient sampling method. However, within the sampling taken, those data points that are not sampled will not have their labels and these data points go by the name of unlabeled or unclustered data points. The problem confronting the investigator is - how does one allocate the unlabeled data point into appropriate clusters [13, 16, 17]. Without loss of generality, the goal of clustering is to allocate every data point into an appropriate cluster. Therefore an efficient algorithm is necessary to allocate the unclustered data points into proper clusters [18]-[20].

In numerical domain [4, 8, 9, 10], the distance between the unclustered data point and the centroid of the cluster give the similarity between them. Each unclustered data point can be assigned to a cluster with the minimal distance. In reality, categorical attributes also prevalently exist in real data. Therefore, allocating unlabeled data point to appropriate clusters remains a challenging issue in the categorical domain considering that the centroid of cluster is difficult to define.

S.Chen proposed Maximal Resemblance Data Labeling (MARDL) mechanism [16] to partially remedy the difficulty. MARDL has two phases: a. cluster analysis phase and b. data labeling phase. In cluster analysis phase, a cluster representative is generated to characterize the clustering result. Chen attached significance to Node Importance Representatives as a categorical cluster representative, emphasizing the importance of attribute values in that cluster [16, 21, 22]. In data labeling phase, an appropriate cluster label was given to unclustered data point according to maximal resemblance, which generates points of similarity based on Node Importance Representative (NIR) and N-Node Importance Representative (NNIR) values. This facilitates allocation to each categorical unclustered data point to the appropriate cluster called data labeling. This paper sets out to investigate and draw comparisons between resemblance values obtained using both NIR and NNIR.

The paper is organized as follows. Section 2 supplies the relevant background to the study; section 3 deals with basic definitions and cluster representatives, such as *N-Node Importance Representative* and *Node Importance Representatives* while section 4 is devoted to cluster analysis, data labeling methods in use for investigation and comparison of NIR and NNIR methods, while the final section – section 5 , concludes the study with recommendations.

## 2  Review of Related Literature

This section provides an exhaustive discussion of various clustering algorithms on categorical data along with cluster representatives and data labeling [23-25]. Cluster representative is used to summarize and characterize the clustering result, which is not discussed in a detailed fashion in categorical domain unlike numerical domain. In K-modes algorithm [4], the most frequent attribute value in each attribute domain of a cluster represents what is known as a *mode* for that cluster. Finding modes may be simple, but only use one attribute value in each attribute domain to represent a cluster is questionable. It composed of the attribute values with high co-occurrence.

ROCK clustering algorithm [26] is a form of agglomerative hierarchical clustering algorithm. This algorithm is based on links between data points, instead of distances between data points. The notion of links between data helps overcome the problems with distance based coefficients. The link between point $i$ $(p_i)$ and point $j$ $(p_j)$, denoted as $link(p_i,p_j)$, and is defined as the number of common neighbours between $p_i$ and $p_j$. ROCK's hierarchical clustering algorithm accepts as input the set S of n sampled points as the representatives of those clusters, drawn randomly from the original data set to be clustered, and the number of desired clusters k. The procedure begins by computing the number of links between pairs of points. The number of links is then used in algorithm to cluster the data set. The first step in implementing the algorithm is to create a Boolean matrix with entries 1 and 0 based on adjacency matrix. The entry is 1 if the two corresponding points are adjacent neighbours or 0 if otherwise. As this algorithm simply focuses on the adjacents of every data point, some data points may be left out or ignored; hence an algorithm based on entropy of the data points is assumed.

In the statistical categorical clustering algorithms [27] such as COOLCAT [21] and LIMBO [28], data points are grouped based on the statistics. In algorithm COOLCAT, data points are separated in such a way that the expected entropy of the whole arrangements is minimized. In algorithm LIMBO, the information bottleneck method is applied to minimize the information lost which resulted from summarizing data points into clusters. However, these algorithms perform clustering based on minimizing or maximizing the statistical objective function, and the clustering representatives in these algorithms are not clearly defined. Therefore, the summarization and characteristic information of the clustering results cannot be obtained using these algorithms [29]. A different approach is called for, which is the aim of the paper.

## 3  N-Node Importance Representative

### 3.1    Notations

Assume that C is a clustering result which consists of C={ $c_1, c_2, \ldots c_k$} where $c_i$, ( i= 1,2, ….k) is the $i^{th}$ cluster. There are $m_i$ data points in cluster $c_i$, whose class label is $C_i^*$. i.e., $c_i = \{p_{(i,1)}, p_{(i,2)}, \ldots p_{(i,m_i)}\}$, where each data point is a vector of q attribute values, i.e., $p_{(i,j)} = \{p_{(i,j)}^1, p_{(i,j)}^2, \ldots p_{(i,j)}^q\}$. Let A={$A_1, A_2, \ldots A_q$}, where $A_a$ is the $a^{th}$ categorical attribute, 1≤a≤q. The unlabeled data set $U = \{p_{(U,1)}, p_{(U,2)}, \ldots p_{(U,m_i)}\}$ is also given, where $p_{(U,j)}$ is the $j^{th}$ data point in data set U. Without loss of generality, the attribute set of U is A. The aim of MARDL is "*to decide the most appropriate cluster label $c_i^*$ for each data point of U*".

We have taken an example of Fig. 1. Here there are three attributes $A_1$, $A_2$ and $A_3$ and three clusters $c_1$, $c_2$ and $c_3$ and unlabeled data set U. The task is to label all the data points of set U of most appropriate cluster. Before assigning values and beginning data labeling, we define the following terms.

**Node:** A *Node* $I_r$ is defined as attribute name + attribute value.

Basically a node is an attribute value, and two or more attribute values of different attributes may be identical, where those attribute domains intersection is non-empty, which is possible in real life. To avoid this ambiguity, we define node not only with attribute value and also with attribute name. For example, Nodes[height=60-69] and [weight=60-69] are different nodes even though the attribute values of attributes height and weight are same i.e.60-69. Because the attribute names height and weight are different then the nodes are different.

**n-nodeset:** An *n-nodeset*, $I_r^n$, is defined as a set of n-nodes in which every node is a member of the distinct attribute $A_a$.

A nodeset is simply a collection of nodes. If there are *n* nodes in that collection then that nodeset is n-nodeset. A 1- nodeset contains only one node. For example {[$A_1$=a]} is a one nodeset. Similarly {[$A_2$=b], [$A_3$=c]} is an example for a 2-nodeset. However, {[$A_1$=a], [$A_1$=b]} is not a 2-nodeset because both attribute values {a} and {b}, are values of same attribute $A_1$.

**Independent nodesets:** Two nodesets $I_r^{n1}$ and $I_r^{n2}$ in a represented cluster are said to *independent* if they do not form larger nodesets and do not contain nodes from the same attributes.

| Cluster $c_1$ | | | Cluster $c_2$ | | | Cluster $c_3$ | | | Unlabeled dataset U | | |
|---|---|---|---|---|---|---|---|---|---|---|---|
| $A_1$ | $A_2$ | $A_3$ | $A_1$ | $A_2$ | $A_3$ | $A_1$ | $A_2$ | $A_3$ | $A_1$ | $A_2$ | $A_3$ |
| a | m | c | c | f | a | c | m | c | a | m | c |
| b | m | b | c | m | a | c | f | b | c | m | a |
| c | f | c | c | f | a | c | m | b | b | f | b |
| a | m | a | a | f | b | b | m | c | a | f | c |
| a | m | c | b | m | a | a | f | a | .. | .. | .. |

Fig. 1 An example of a data set with three clusters and Unlabeled dataset.

The above definition indicates that two node-sets $I_r^{n1}$ and $I_r^{n2}$ together $(i.e., I_r^{n_1} \cap I_r^{n_2})$ do not come in the cluster representative and nodes in $I_r^{n1}$ and $I_r^{n2}$ do not come from the same attribute. If the two node-sets are independent, then the probability of their intersection in the cluster can be estimated by multiplying the probabilities of the two node-sets in question.

3.2　　Node Importance Representative

　　NIR is used to represent a cluster as the distribution of the attribute values. A node $I_r$, is defined as attribute name plus attribute value. NIR considers both the intracluster and intercluster similarity. The importance of the node in a cluster is measured making use of the two concepts that figure below:

(i) The node is important in the cluster when the frequency of the node is high in this cluster.

(ii) The node is important in the cluster if the node appears predominantly in this cluster rather than in other clusters.

The idea of NNIR is to represent a cluster as the distribution of the n-nodesets, which are already defined in this section. NNIR is an extension **of** NIR where each attribute value combinations are considered to characterize the clustering results.

Based on the above two concepts, we define the n-nodeset $I^n_{ir}$ in equation (1):

$$w(c_i, I^n_{ir}) = \frac{|I^n_{ir}|}{m_i} * f(I^n_r) \tag{1}$$

$$f(I^n_r) = 1 - \frac{-1}{\log k} * \sum_{y=1}^{k} p(I^n_{yr}) \log(p(I^n_{yr}))$$

Where

$$p(I^n_{yr}) = \frac{|I^n_{yr}|}{\sum_{z=1}^{k^t} |I^m_{zr}|}$$

Where $m_i$ is the number of data points in cluster $C_i$, $|I^n_{ir}|$ is the frequency of the nodeset $I^n_{ir}$, and $k$ is number of clusters, since this is a product of two factors. The first factor is the probability of $I_{ir}$ being in $C_i$ using rule (i), which aims to maximize the intra cluster similarity and the second factor is the weighting function arrived at using rule (ii) which minimizes the inter cluster similarity. Entropy E(x) is defined as $E(X) = \sum_{y=1}^{k} p(I_{yr}) \log(p(I_{yr}))$, a measurement of information and uncertainty on a random variable [30]. The minimum entropy value of a node between clustered is equals to *logk*. The entropy value of a node between clusters is divided by *logk* to normalize the weighting function from zero to one. If we subtract this quantity by one, the node containing large entropy will obtain a small weight. Since the range probability of $I^n_{ir}$ being in $c_i$ is zero to one, it is implied that $W(c_i, I^n_{ir})$ range is also [0 1].

## 4  Cluster Analysis and Data Labeling Methods and Comparison of NIR/NNIR results

In cluster analysis phase, an NNIR lattice tree is constructed by considering all the combinations of attribute values which occur in the cluster with their n-nodeset values. This helps in establishing the tree which represents the clustering results. Because the size of the tree is large as the number of attribute value combinations is much, NNIR tree pruning algorithms (i.e. Threshold Pruning, Relative Maximum Pruning and Hybrid Pruning, discussed in [16]) are applied so as to preserve significant n-nodesets and ignore insignificant node-sets.

In data labeling phases, a lot of resemblance is found between unlabeled data point and the existing clusters; using MARDL, the cluster label $c^*$ pertaining to the relevant unlabeled data point is identified.

Nodeset combination: For a given cluster $c_i$, having a fixed NNIR tree, and an unlabelled data point $p_{(U, J)}$, the nodeset combination is defined by a set of nodesets whose union is $p_{(U, J)}$ and are independent of each other. These are also found in the NNIR tree of $c_i$.

For example, the nodeset combinations of unlabeled data point $p_{(U, 1)}$ ={[$A_1$=a], [$A_2$=m], [$A_3$=c]} given in Fig.1 are the following :

{[$A_1$=a], [$A_2$=m]}, {[$A_3$=c]}

{[$A_1$=a], [$A_3$=c]}, {[$A_2$=m]}

{[$A_1$=a]}, {[$A_2$=m]}, {[$A_3$=c]}

**Resemblance:** Suppose a cluster $c_i$ is represented by an NNIR tree and a given unlabeled data point $p_{(U, j)}$, then the formula (2) gives the resemblance between these two.

$$R(c_i, p_{(U,j)}) = max \prod_u \frac{|I^{n_u}_{ir_u}|}{m_i} * E(f(I^{n_u}_{ir_u}))$$

(2)

Where $0 < n_u = n$, $n_u = n$  $\forall I_{in_u}^{n_n}$ are independent to each other and there union is $p_{(U, j)}$.

The resemblance with all $n_u$- nodesets are thus found. The combination which gives maximum resemblance is chosen as the resemblance $R(c_i, p_{(U, J)})$, between $p_{(U, j)}$ and $c_i$. Since all $n_u$- nodesets are independent of each other, the probability of the combination in cluster can be measured by the product of the probabilities of $I_{iru}^{nu}$ in cluster $c_i$, and the weight of the combination is estimated by the expected value of the weight of $I_{in_u}^{n_n}$ i.e. $E(f(I_{in_u}^{n_n}))$.

**Example 1:** A dataset given in Fig.1 with three different attributes $A_1$, $A_2$ and $A_3$ and 15 data points which are divided into three clusters $c_1$, $c_2$ and $c_3$ using some clustering technique. In Fig.1 an unlabeled dataset U is also given. The following lattice tree in Fig. 2 is the NNIR tree of cluster $c_1$ and similarly the lattice NNIR trees of cluster $c_2$ and $c_3$ are also given in Fig. 3 and Fig. 4 respectively.

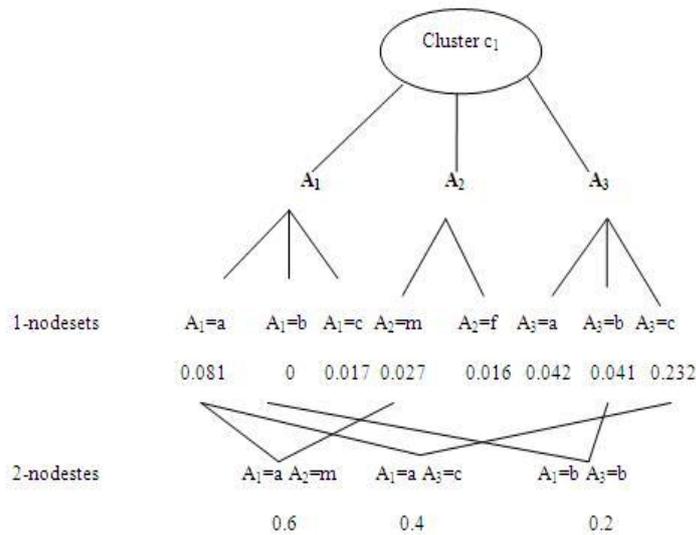

Fig.2 The pruned NNIR tree of Cluster $c_1$

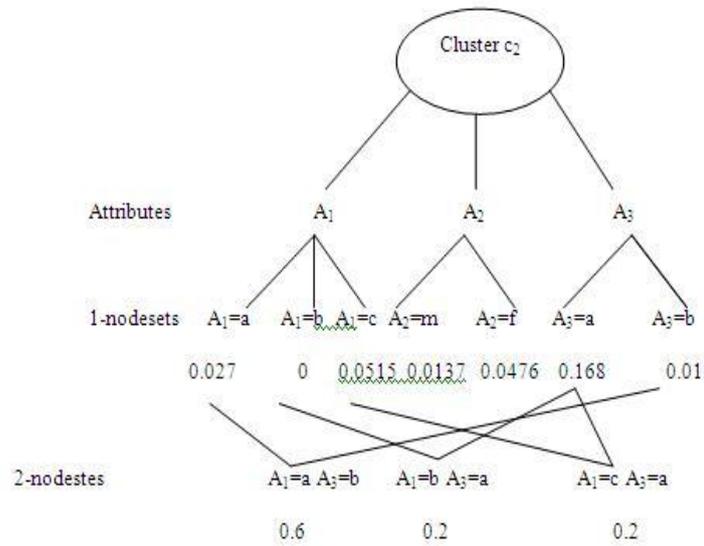

Fig.3 The pruned NNIR tree of Cluster $c_2$.

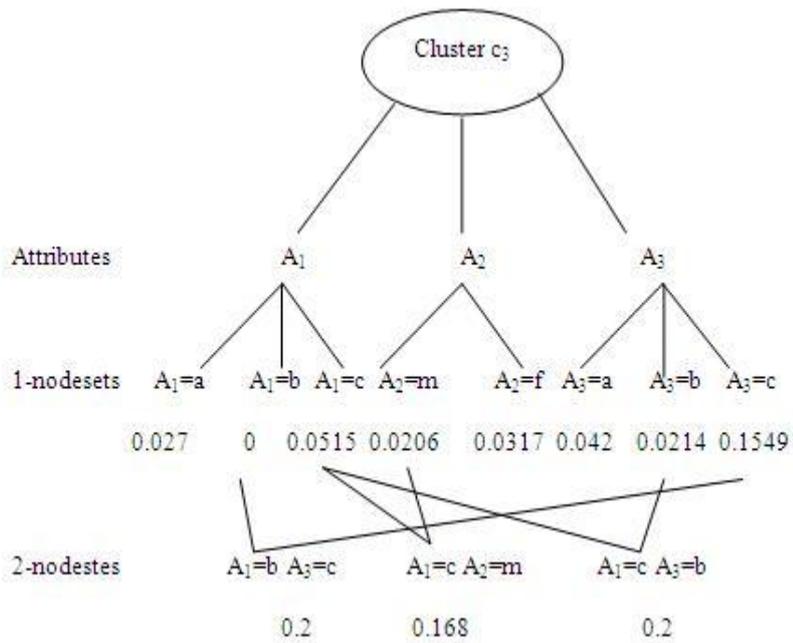

Fig.4 The pruned NNIR tree of Cluster $c_3$

Suppose an unlabeled data point (b, f, b) i.e. $p_{(U, 3)}=\{[A_1=b], [A_2=f], [A_3=b]\}$ is taken, then

The nodeset combination of $p_{(U, 3)}$ are the following.

(i)     $\{[A_1=b], [A_2=f]\}, \{[A_3=b]\}$

(ii)    $\{[A_1=b], [A_3=b]\}, \{[A_2=f]\}$

(iii)   $\{[A_1=b]\}, \{[A_2=f], [A_3=b]\}$

(iv)    $\{[A_1=b]\}, \{[A_2=f]\}, \{[A_3=b]\}$

Out of the above four nodeset combinations of $p_{(U, 3)}$, (ii) and (iv) is present in NNIR tree of $c_1$, as given in Fig.2. The resemblance of these two nodeset combinations with cluster $c_1$ using equation (2) is evident.

Therefore the resemblance with $\{[A_1=b], [A_3=b]\}, \{[A_2=f]\}$

= (1/5)*(1/5)*((2/3)*1+(1/3)*0.07948))=0.027,

And the resemblance with $\{[A_1=b]\}, \{[A_2=f]\}, \{[A_3=b]\}$

= (1/5)*(1/5)*(1/5)*((1/3)*0+(1/3)*0.07948+(1/3)*0.0536)) =0.00035.

The combination (ii) gives the maximum resemblance value with cluster $c_1$ then $R(c_1, p_{(U, 3)})=0.0277$

Similarly, we find resemblance of the same nodeset $p_{(U, 3)}$ with cluster $c_2$ as follows. Out of the above four nodeset combinations of $p_{(U, 3)}$ nodesets (iii) and (iv) are present in NNIR tree of $c_2$ given in Fig.3.The resemblance of these two nodeset combinations with cluster $c_2$ is established using equation (2).

Therefore the resemblance with $\{[A_1=b]\}, \{[A_2=f]\}, [A_3=b]\}$

= (1/5)*(1/5)*((1/3)*0+(2/3)*0.3690))=0.00984

And the resemblance with $\{[A_1=b]\}, \{[A_2=f]\}, \{[A_3=b]\}$

= (1/5)*(3/5)*(1/5)*((1/3)*0+(1/3)*0.07948+(1/3)*0.0536))=0.00106

Combination (ii) gives the maximum resemblance value with cluster $c_2$ then $R(c_2, p_{(U, 3)})=0.00984$.

Therefore, with these two clusters, we find maximum resemblance of $p_{(U, 3)}$ with cluster $c_1$ ; cluster $c_1$ is therefore most appropriate for the unlabeled data point $p_{(U, 3)}$ according to NNIR method. But according to the method introduced by S.Chen in [13], this unlabeled data point $p_{(U, 3)}$ cluster label is found as $c_2^*$. i.e. the most appropriate cluster of this unlabeled data point $p_{(U, 3)}$ is cluster $c_2$, as shown below.

In S.Chen's method, not all the nodeset combinations of unlabeled data point are considered; only a single node importance (i.e. NIR) is considered in MARDL method to decide the cluster label and that gives rise to the discrepancy.

Now consider the same clusters $c_1$, $c_2$ and $c_3$ as shown in Fig. 1, and the unlabeled dataset U. Consider same unlabeled data point $p_{(U,\ 3)}$={[$A_1$=b], [$A_2$=f], [$A_3$=b]}

Using Chen's method one obtains resemblance, using the formula (3):

$$R(p_j, c_i) = \sum w(c_i, I_{ir}) \qquad (3)$$

Since we measure the similarity between the data point and cluster $c_i$ as $R(p_j, c_i)$, the cluster with the maximal resemblance is the most appropriate cluster for that data point.

$$R(c_1, p_{(U,\ 3)}) = 0 + 0.015 + 0.0107 = 0.0257$$
$$R(c_2, p_{(U,\ 3)}) = 0 + 0.047 + 0.01 = 0.057$$

But the maximum value is obtained in $c_2$ with this method thereby contradicting the result of the method used by us to arrive the result.

The new method is therefore advanced to remedy and better the results obtained using Chen's method. The new formula for finding resemblance which can be used in MARDL clustering given in equation (4):

$$R(I_{iru}^{nu}, c_i) = \max_{nu} \sum E(w(I_{iru}^{nu}, c_i)) \qquad (4)$$

Using of this formula in MARDL, both the methods yield the same results.

## 5. Conclusions

This paper employed MARDL method to allocate each unlabeled data point for an appropriate cluster because in sampling techniques, clustering is done on a small sampled data set from the categorical database. This is because sampling technique clustering uses many unlabeled data points to which appropriate cluster labels should be given. This MARDL method works based on NIR/NNIR. NNIR is an extension of and an improvement on NIR which works better than NIR because all nodeset combinations are considered to find the resemblance. Both NIR and NNIR have been proposed and used in MARDL by S.Chen. This paper investigated and compared the results derived using NIR and NNIR for cluster and unlabelled data points and found that the results obtained are contradictory, revealing different values for different methods when they ought to be the same. To redress this, another method was proposed to find the maximal resemblance between an unlabeled data point and a cluster.

**Acknowledgments.** We wish to thank our supervisor Dr.Vinay Babu, JNTUniversity Hyderabad, India and all our friends who helped us in writing this paper.

**International Journal of Computer Engineering Science (IJCES)**
Volume 2 Issue 8 (August 2012)      ISSN : 2250:3439
https://sites.google.com/site/ijcesjournal      http://www.ijces.com/19. Venkateswara Reddy.H, Viswanadha Raju.S, Our-NIR: Node Importance Representative for Clustering of Categorical Data, *International Journal of Computer Science and Technology* , pp. 80-82,2011.

20. Venkateswara Reddy.H, Viswanadha Raju.S, POur-NIR: Modified Node Importance Representative for Clustering of Categorical Data, *International Journal of Computer Science and Information Security*, pp.146-150, 2011.

21. Barbara, D., Li, Y. and Couto, J.,  Coolcat: An Entropy-Based Algorithm for Categorical Clustering, *ACM International Conf. Information and Knowledge Management (CIKM)*, 2002.

22. Venkateswara Reddy.H, Viswanadha Raju.S, A Threshold for clustering Concept – Drifting Categorical Data, *IEEE Computer Society, ICMLC* 2011.

23. Ganti, V., Gehrke, J. and Ramakrishnan, R, CACTUS—Clustering Categorical Data Using Summaries,  *ACM SIGKDD,* 1999.

24. Gibson, D., Kleinberg, J.M. and Raghavan,P. Clustering Categorical Data An Approach Based on Dynamical Systems, *VLDB* pp. 3-4, pp. 222-236, 2000.

25. Vapnik, V.N, *The nature of statistical learning theory,*(Springer,1995).

26. Guha,S., Rastogi,R. and Shim, K, ROCK: A Robust Clustering Algorithm for Categorical Attributes, *International Conference On  Data Eng. (ICDE)*, 1999.

27. Sudipto Guha, Adam Meyerson, Nina Mishra, Rajeev Motwani, and Liadan O'Callaghan, Clustering data streams: Theory and practice, *IEEE Transactions on Knowledge and Data Engineering*, pp.515–528, 2003.

28. Andritsos, P, Tsaparas, P, Miller R.J and Sevcik, K.C.Limbo: Scalable Clustering of Categorical Data, *Extending Database Technology (EDBT)*, 2004.

29. Shannon, C.E, A Mathematical Theory of Communication, *Bell System Technical J.*, 1948.

30. Gluck, M.A.  and Corter, J.E., Information Uncertainty and the Utility of Categories, *Cognitive Science Society,* pp. 283-287, 1985.
15